\documentclass[aps,pra,floatfix,showpacs,noshowkeys,epsfig,graphics]{revtex4-1}

\usepackage{graphicx}
\usepackage{amsmath}
\usepackage{amsfonts}
\usepackage{amssymb}

\newcommand{\newc}{\newcommand}
\newc{\beq}    {\begin{equation}}
\newc{\eeq}    {\end{equation}}

\newc{\beqa}    {\begin{eqnarray}}
\newc{\eeqa}    {\end{eqnarray}}
\newc{\bs}    {\section}
\newc{\no}    {\\ \nonumber}

\newtheorem{theorem}{Theorem}

\newc{\st}    {\stackrel}

\begin{document}
\title{ Physics from information }
\author{Jae-Weon Lee}\email{scikid@gmail.com}
\affiliation{ Department of energy resources development,
Jungwon
 University,  5 dongburi, Goesan-eup, Goesan-gun Chungbuk Korea
367-805}

\date{\today}

\begin{abstract}
This is an ongoing review on the idea that  the phase space information loss at causal horizons
is the key ingredient of major physical laws.
 Assuming  that information is fundamental
and the information propagates with finite velocity,
one can find that basic physical laws such as Newton's second law and Einstein's equation
 simply describe the energy-information relation (dE=TdS) for  matter or space time crossing
causal  horizons.
Quantum mechanics  is related to the  phase space information loss  of matter crossing the Rindler horizon,
which explains why superluminal communication is impossible even with quantum entanglement.
This approach also explains the origin of  Jacobson's thermodynamic formalism of  Einstein gravity and Verlinde's entropic gravity.
When applied to a cosmic causal horizon, the conjecture can reproduce  the observed dark energy
and the zero cosmological constant.
Quantum entanglement, path integral, and holography are also natural consequences of this theory.
 \end{abstract}

\maketitle
\section{Introduction}

For thousands  of years great minds of mankind  tried to  find the most fundamental element
in nature such as four elements,  atoms, quarks, and strings.
Nowadays, there is  a  hope   that this quest will eventually  lead us to  a `theory of everything'
 reconciling general relativity with  quantum mechanics and other forces.
In this perspective,   configurations of a  fundamental object such as a superstring
should represent all known particles and their species.
However,  to have configurations, the object should have some internal structure and this implies that the object should
consist of even smaller objects. This brings us an obvious logical paradox.

On the other hand, there is a long history of the conception that the universe  is actually made of
abstract entities like logic rather than material objects.
 A famous example is
 the Pythagorean who believed  that numbers are fundamental constituents of the nature.
Interestingly,
recent developments of quantum information
science revealed that abstract information can play
a fundamental roll in the physical world.
This idea can be represented  by an implicative  slogan in quantum information community, ``It from
Bit!''

There are many observations supporting the slogan.
For instance, it was shown that
quantum mechanics and special
relativity miraculously cooperate so as not to allow
super-luminal information transfer (See for example ~\cite{PhysRevA.78.022335}),
 and this no-signalling condition could be a basic principle of physics.
Furthermore, Landauer's principle~\cite{landauer} stating that  erasing information requires energy consumption
implies an intrinsic relation between information and energy.
It was also suggested that wavefunctions in quantum mechanics actually represent
information of a system~\cite{zeilinger1999}  or relations~\cite{Rovelli:1995fv} rather than particles or waves.

Studies of black hole physics after  Bekenstein and Hawking  have
consistently implied that there is a deep connection among gravity, thermodynamics and information~\cite{Bekenstein}.
Recently proposed Verlinde's
 idea~\cite{Verlinde:2010hp} linking  gravity to entropic force
 enhance this viewpoint,
 because entropy can be interpreted as a measure of information. He derived  Newton's second law and  Einstein's equation from the relation
between the number of degrees of freedom $N$ of a holographic screen and energy $E\sim N T$ in a volume enclosed by  the screen.
Here, $T$  is the temperature  of the screen.
Padmanabhan ~\cite{Padmanabhan:2009kr} also  proposed that
 classical gravity can be derived from the equipartition energy of horizons.
 These works, influenced by Jacobson's proposal that Einstein equation describes
the first law of thermodynamics at  local Rindler horizons, attracted much interest in community
~\cite{Zhao:2010qw,Cai:2010sz,Cai:2010hk,Myung:2010jv,Liu:2010na,Tian:2010uy,Pesci:2010un,Diego:2010ju,
Vancea:2010vf,Konoplya:2010ak,Culetu:2010ua,Zhao:2010vt,Ghosh:2010hz,Munkhammar:2010rg,Kuang:2010gs}.
All these works emphasize mainly the connection between thermodynamics and gravity.

On the other hand, in a series of works \cite{myDE}, the author and co-workers (LLK hereafter) suggested  that
information plays a crucial role
in gravitational systems such as dark energy and black holes.
For example, in 2007  LLK  presented a new idea that
a cosmic
causal horizon with a radius $r\sim O(1/H)$ could have
  Hawking temperature $T_h\propto 1/r$,  quantum information theoretic  entropy
$S_h\propto r^2$ represented by bits,
and hence, a kind of  thermal
energy $E_{h}\sim T_h S_h\propto r$, which can be identified to be the dark energy with
density $\rho_{h}\sim  r^{-2}\sim O(M_P^2 H^2)$. Here, $M_P$ is the Planck mass and $H$ is the Hubble parameter.
We set the Boltzman constant $k_B=1$.
This energy
corresponds to the quantum vacuum energy of a spatial region
bounded by the horizon and is related to information erasing process due to the expanding cosmic horizon
and also possibly to quantum entanglement of the vacuum.
LLK also suggested that a black hole mass has a similar quantum information theoretic origin,
and that Jacobson's formalism about Einstein gravity actually represents
information loss process at local Rindler horizons in a curved spacetime.

Since  entropy is usually proportional to $N$, there is a clear similarity between this informational energy $E_h$ and the equipartition
 energy $E$ considered by Verlinde. However, there are also some differences between two approaches which will be shown below.

In this paper,  based on these works, it is suggested that major physics such as
quantum mechanics, Einstein gravity and Newton's mechanics
are  simply describing  information processing at causal horizons.

 \section{It from Bit}

Let me start by summarizing some well-known physical principles and laws.
\begin{enumerate}
\item
Landauer's principle: To erase information $dS$, at least energy $dE= TdS$ should be consumed. \\
     $\rightarrow$ Information is related to thermal Energy
\item
$E=m c^2$: Energy is related to Mass (matter)
\item
 Einstein Equation, $G_{\mu\nu}=8\pi G ~T_{\mu\nu}$
  $\rightarrow$ Matter generates Gravity
 \item
 Unruh effect: Quantum fluctuation  looks thermal to some observers
\end{enumerate}
Now, in a very naive language, my observations can be summarized as follows. By combining
the principles 1 and 2, one can see
``Matter is related to information".
On the other hand, $1+2+3$ implies  ``Gravity is related to information".
1+4 means ``Quantum mechanics is related to information".
Although these  propositions should be justified by more rigorous  reasonings,
this brief argument shows the essence of the idea.

Inspired by the above principles and laws, it is now suggested that we can choose the followings
as   new and
general guiding principles which any physical law is based on.\\

$\bullet$ Guiding principles

\begin{enumerate}
{\it
\item  General equivalence:
All systems of reference (coordinates) are equivalent for formulating physical laws regardless of their motions.
\item Information  has a finite density and speed; the quantity of information contained in a finite object is finite, and there is a maximal speed of classical information propagation, namely, the light velocity.
\item  Information is fundamental:
 Physical laws regarding an object (matter or spacetime) should be such that  they
 respect  observers' information about the object.
 }
\end{enumerate}


Note first that we are  assuming neither quantum mechanics nor Einstein gravity.
They  shall be derived from above assumptions.
We need to assume  the metric nature of spacetime
and ignore any fluctuation of spacetime.

\begin{figure}[tpbh]
\includegraphics[width=0.3\textwidth]{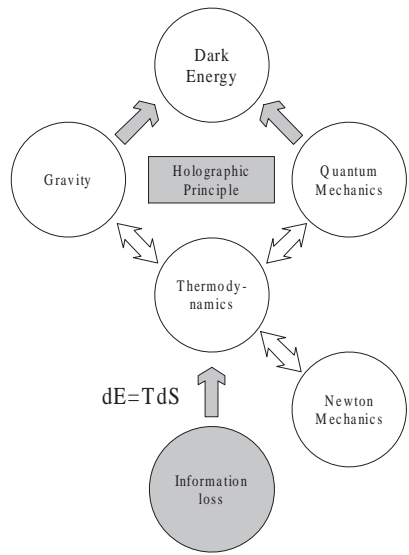}
\caption{
Relation between various physics fields.
Information seems to be the root of all physics.
Causal horizons for some observers act as an information barrier, and thermodynamics occurs as a result.
Then, an information-energy (i.e., entropy-energy) relation  $dE=TdS$ at the horizons
 lead to main equations of physics.
}
\end{figure}

Some of these assumptions deserve more explanation.
Finite information propagation velocity implies that there is an information
barrier in spacetime for some observers.
This barrier could be, for example, a Rindler horizon, a black hole event horizon or  a light cone.
Thus, there could be a situation where matter (particles or waves)
crosses the causal horizon for an observer.
Then, the observer can get no more information about the phase space (position and velocity) of the matter.
It is reasonable
that this  ignorance of the observer about the matter should be represented
by the increase of the information theoretic entropy $S$ (for example Shannon entropy or entanglement
 entropy) of the horizon.
According to the holographic principle this should be accompanied by
the horizon area increase.
Furthermore, due to Landauer's principle or the second law of thermodynamics, there should be
some kind of `thermal energy' $dE= TdS$. That means the usual first law of thermodynamics in gravitational systems
 is actually the second law disguised.
Major physical laws such as Einstein equation and Newton's equation
seem to simply represent this information-energy relation.
\\

In short I suggest the following conjecture.
\\
\\
$\bullet$ Conjecture:
{\it Main physical laws
 simply describe the phase space information loss of  matter or space time crossing a
causal  horizon for an observer}
\\

From this conjecture, the following results are derived (See Fig. 1.).
Quantum mechanics arises from ignorance of observer outside of a causal horizon  about matter inside the horizon
(section III).
For an accelerating Rindler observer relative to a  particle this leads to Newton's second law as in  Verlinde's formalism
(section IV).
For a local inertial frame in curved spacetime the conjecture leads to Einstein equation through Jacobson's formalism
and naturally explain the origin of Verlinde's entropic gravity (section V).
This theory also explains the origin of holography and quantum entanglement (section VI).
Finally, if we apply the conjecture to a cosmic causal horizon, we obtain dark energy comparable to observed one
and zero cosmological constant (section VII).

\section{Quantum Mechanics from information loss}

In this section it will be shown that quantum field theory (QFT), and hence quantum physics, is not fundamental and can be derived by considering
phase space  information loss
of matter crossing Rindler horizons ~\cite{fopleejw}.
Let us begin by considering an
accelerating Rindler observer $\Theta_R$ with acceleration $a$ in $x_1$ direction
 in a flat spacetime with coordinates $X=(t,x_1,x_2,x_3)$ (See Fig. 1).
The Rindler coordinates $\xi=(\eta,r,x_2,x_3)$ for the observer are defined with
\beq
\label{rindler}
ct= r~ sinh (a \eta/c),~ x_1= r ~cosh (a \eta/c)
\eeq
 on the Rindler wedges.
There is an inertial Minkowski observer $\Theta_M$ too.
Now, consider a field $\phi$
 flowing across  the Rindler horizon  at a point $P$
  and entering the future wedge $F$. A configuration for
 $\phi(x,t)$ is not necessarily meant to be classical but to be just some physical function of spacetime.
 It is important to note that in this theory
  the field $\phi$ cannot have a specific value before measurements according to our assumptions,
unless the relevant observer gets the  information about the field value in advance.

\begin{figure}[tpbh]
\includegraphics[width=0.3\textwidth]{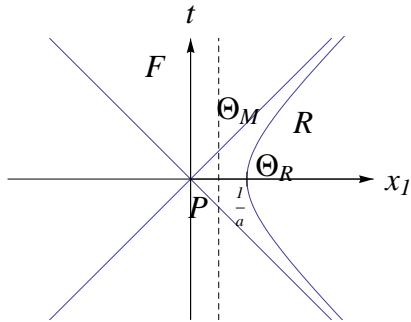}
\caption{ Rindler chart for  the observer $\Theta_R$ (curved line),
who has no accessible information about  field $\phi$
in  a causally disconnected  region $F$.
Thus, the observer can only estimate a probabilistic distribution of the field,
which turns out to be thermal and
equal to that of a quantum field for inertial observer $\Theta_M$ (dashed line) in Minkowski spacetime.
}
\end{figure}

As the field enters  the Rinder horizon for the observer $\Theta_R$,
 the observer shall not get  information about phase space information (configurations and momentums)  of $\phi$  any more
and  all what the observer can expect about $\phi$ evolution beyond the horizon
 is a probabilistic distribution $P[\phi]$ of $\phi$ beyond the horizon.
Already known information about $\phi$ acts as  constraints for the distribution.
I suggested that this ignorance is the origin of quantum randomness.
Physics in the F wedge should reflect the ignorance of the observer in the R wedge,
if information is fundamental~\cite{fopleejw}.

One constraint  comes from the energy conservation
\beq
\sum_{i=1}^n  P[\phi_i]H(\phi_i) = E,
\eeq
where $H(\phi_i)$ is the Hamiltonian as a function of the i-th configuration
of the field $\phi_i$ and $E$ is its expectation.
Another  one is the unity of  the probabilities
$
\sum_{i=1}^n  P[\phi_i] = 1.
$
Then, using  Boltzmann's  principle of maximum entropy  one can  calculate the probability distribution
 estimated by the Rindler observer
\beq
\label{P}
P[\phi_i] = \frac{1}{Z} \exp\left[- \beta H(\phi_i)\right],
\eeq
where $\beta$ is the Lagrangian multiplier, and the partition function is
\beq
\label{Z}
 Z = \sum_{i=1}^n  \exp\left[- {\beta H(\phi_i)} \right]=tr~ e^{-\beta H},
 \eeq
 where the trace is assumed to be performed with a (classical) discrete vector basis.
 Lisi showed a related derivation of the partition function
by assuming a universal action reservoir for information~\cite{lisi-2006}.

 From now on, let us consider  a continuum limit for
 a scalar field $\phi$  with Hamiltonian
 \beq
 H(\phi)=\int d^3 x \left[ \frac{1}{2}\left( \frac{\partial \phi}{\partial t}\right)^2+
 \frac{1}{2}\left( {\nabla \phi}\right)^2 + V(\phi) \right]
 \eeq
 and  potential $V$.
  For the Rindler observer with the coordinates $(\eta,r,x_2,x_3)$  the proper time variance is
  $ard\eta$ and hence the Hamiltonian is changed to
  \beq
 H_R =  \int_{r>0} dr dx_\bot~ ar
 [
  \frac{1}{2}\left( \frac{\partial \phi}{ar \partial \eta}\right)^2
 +
 \frac{1}{2}
 \left(\frac{\partial \phi}{ \partial r} \right)^2+\frac{1}{2}
 \left( {\nabla_\bot \phi}\right)^2
 +  V(\phi)] ,
 \eeq
where $\bot$ denotes the plane orthogonal to $(\eta,r)$ plane.
Then, Eq. (\ref{Z}) becomes Eq. (2.5) of Ref. ~\cite{1984PhRvD..29.1656U};
\beq
\label{Z_R}
 Z_R = tr~ e^{-\beta H_R}.
 \eeq
It is important to notice
  that $Z$ (and hence $Z_R$) here is not a quantum partition function but a
  statistical partition function corresponding to the uncertain
 field configurations  beyond the horizon.

Unruh showed  ~\cite{1984PhRvD..29.1656U} that the
real-time thermal Green's functions of the Rindler observer with $Z_R$ are equivalent to
the vacuum Green's function in Minkowski coordinates.
Thus, as well known, the Minkowski vacuum is equivalent to thermal states for the Rindler observers.
What is newly shown in Ref. ~\cite{fopleejw} is that the  thermal partition function $Z_R$
assumed in  Ref. ~\cite{Crispino:2007eb}
 is actually from the phase space  information loss  beyond the Rindler horizon
and, therefore, the QFT formalism is equivalent to  the purely information
theoretic formalism.
Recall that Eq. (\ref{Z_R}) was derived without using any quantum physics.
Since quantum mechanics can be thought to be a single particle approximation of QFT, this implies also
that particle quantum mechanics emerges from information theory applied to Rindler horizons and
is not fundamental.
Another important point here is that thermal nature  of quantum field is due to
the information loss and can be treated as more fundamental than the quantum nature.

This theory explains why superluminal communication is impossible even using quantum nonlocality (entanglement).
Quantum randomness and hence quantum correlation originates from the very fact that information cannot be
 sent faster than light.

In Ref.~\cite{Crispino:2007eb} it was shown by analytical continuation
that in the Rindler coordinates $Z_R$ is $mathematically$ equivalent to
 \beq
 Z_R =
   N_0\int_{\phi(0)=\phi(\beta')}
  D\phi ~exp\{-\alpha\int_0^{\beta'} d\tilde{\eta} \int_{r>0} dr dx_\bot~ ar
   \left[
  \frac{1}{2}\left( \frac{\partial \phi}{ar \partial \tilde{\eta}}\right)^2
 +  \frac{1}{2}
 \left(\frac{\partial \phi}{ \partial r} \right)^2+\frac{1}{2}
 \left( {\nabla_\bot \phi}\right)^2
 +  V(\phi)\right]\},
 \eeq
 where we  introduced a constant  $\alpha$  having a dimension
 of  $1/H_Rt$ and $\beta\equiv \alpha \beta'$.

  By further  changing integration variables as
$\tilde{r}=r cos(a\tilde{\eta}),  \tilde{t}= rsin(a\tilde{\eta})$ and choosing $\beta'=2\pi/a\equiv 1/\alpha  T_U$
the  region of integration is transformed from $0\le \tilde{\eta} \le \beta', 0\le r \le \infty$
into the full two dimensional $\tilde{t}- \tilde{r}$  space.
This   $\beta'$ value leads to Unruh temperature
$ T_U={ a}/{2\alpha \pi  }$.
From the well-known QFT result,  one can find $1/\alpha=\hbar$.
This means that the Planck constant $\hbar$ is some fundamental temperature given by nature.

Then, the partition function becomes
\beq
 Z^E_Q =  N_1\int
  D\phi ~exp\left\{-  \frac{I_E}{\hbar}\right\}.
 \eeq
where $I_E$ is the Euclidean action for the scalar field in the inertial frame.
By analytic continuation $\tilde{t}\rightarrow it$,
one can see $Z^E_Q$ becomes  the usual zero temperature quantum mechanical partition function $Z_Q$ for $\phi$.
Since both of $Z_R$ and $Z_Q$ can be obtained from $Z^E_Q$ by analytic continuation, they are
physically equivalent as pointed out in Ref. ~\cite{Crispino:2007eb}.

It is straightforward to extend the previous analysis to quantum mechanics for point particles.
We can imagine a point particle  at a point $P$
just crossing the Rindler horizon and entering the future wedge $F$.
The maximal ignorance of the observer about the particle
 is represented by probability distribution $P[x_i(t)]$
for the i-th possible path that the particle may take.
Then, the partition function is
\beq
\label{ZR}
 Z_R = \sum_{i=1}^n  \exp\left[- {\beta H(x_i)} \right]=tr~ e^{-\beta H},
 \eeq
 where $H$ is the point particle Hamiltonian now.
 Since the usual point particle quantum mechanics
 is a non-relativistic and single particle limit of the quantum field theory,
 we expect $Z_R$ is equal to the quantum partition function for the particle
 with mass $m$  in Minkowski spacetime
 \beqa
 Z_Q &=&
  N_2\int  Dx \exp\left[-\frac{i}{\hbar}\int d\tilde{t}  \left\{\frac{m}{2}
 \left( \frac{\partial x}{\partial \tilde{t}}\right)^2 -  V(x)\right\}  \right]\no
 &=&N_1\int
  Dx ~exp\left\{- \frac{i}{\hbar} I(x_i)\right\},
\eeqa
 where $I$ is the action for the point particle. Then,
as is well known one can associate each path $x_i$ with a wave function $\psi\sim e^{-iI}$,
which leads to  Schr\"{o}dinger equation for $\psi$~\cite{derbes:881}.
Therefore, our theory explains naturally the origin of path integral and
the similarity between quantum mechanical formalism and statistical physics.

 \section{Newton mechanics from information loss}
 Quantum mechanics of the previous section, of course, leads to classical Newton
  mechanics for an appropriate limit ($\hbar \rightarrow \infty$ and $c \rightarrow \infty$ ).
 Alternatively, one can also  directly derive
 Newton mechanics from the information-energy relation based on the partition function
 as in Verlinde's approach. If our theory is sound, two approaches should give a
 same description.

The   free energy $G$ from the partition function of the previous section can be expressed as
\beq
G=-\frac{1}{\beta} ln Z_R.
\eeq
The classical path $x_{cl}$ for the particle  corresponds to the saddle point
 ($Z_R\sim exp[-\beta I_E(x_{cl})$] )~\cite{Banerjee:2010yd},
 where $I_E(x_{cl})$ is the Euclidean action for classical path satisfying the Lagrange equation.
In this limit the free energy becomes
\beq
G\simeq G_{cl}= -\frac{1}{\beta} \left(-\beta I_E(x_{cl})\right)+C=I_E(x_{cl})+ C,
\eeq
where $C$ is a constant.
Since the maximum entropy is achieved when $G$ is minimized, we see that
classical physics with the minimum action corresponds to a maximum entropy condition.
In other words,
the classical path is the typical path  maximizing the Shannon entropy $h[P]$
regarding the phase space information with the constraints for the Rindler observer.

Therefore, one can find that the entropy associated with the thermodynamical interpretation of mechanics and gravity is
related to information of matter crossing the horizon.
For fixed temperature, pressure and volume, the minimum free energy condition $dG=0$ is equivalent to
$dE-TdS=0$, i.e., the first law of thermodynamics or the information-energy relation, $dE=TdS$.
This explains why classical physics can be obtained from thermodynamics as in Verlinde's approach.
The maximum entropy proposal in Verlinde's theory can be easily explained in this theory.

To be concrete,  consider an accelerating test point particle  with acceleration $a$ and mass $m$ (Fig. 1)
and an  observer $\Theta_R$ at rest at the instantaneous distance $\Delta x$ from the particle.
If we accept the general principle of relativity stating that
all systems of reference are equivalent regardless of their motions,
 we can imagine an equivalent situation where the particle is at rest
and the observer $\Theta_R$ accelerates in the opposite direction with acceleration $-a$.

The key idea  is that
for an accelerating object there is always such an
 observer that  the object seems to cross a Rindler horizon of the observer.
 For this observer there is the phase space  information loss,
 and hence, some thermal energy associated.
\begin{figure}[tpbh]
\includegraphics[width=0.3\textwidth]{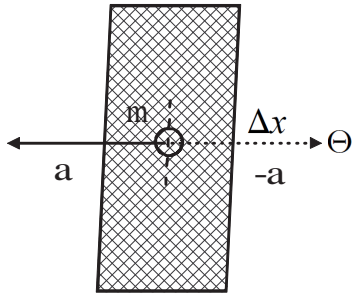}
\caption{
A test particle with mass $m$ is accelerating with acceleration $a$ with respect to
an observer instantaneously  resting at $\Delta x$ from the particle.
Alternatively we can imagine that the particle is at rest
and the observer moves  in the opposite direction with acceleration $-a$.
The observer could see the particle crossing a Rindler horizon (the shaded plane).
}
\end{figure}
If the observer is at a specific distance $\Delta x=c^2/a$,
the observer could see the particle just crossing  his or her Rindler horizon.
Then, the Rindler  horizon hides the information of the particle
and this leads to information loss, which should be compensated by an increase of the
entropy  $S_h$ of the Rindler horizon.
 This distance $\Delta x=c^2/a$
is special, because,  for the observer there, $\tau$ becomes a proper time and
 the Rindler Hamiltonian becomes a physical one
generating $\tau$ translation.
Since the horizon is a Rindler horizon,
we can safely use the Unruh temperature
\beq
\label{unruh}
 T_U=\frac{\hbar a}{2\pi c }
 \eeq
 for the horizon.

Then, if we accept the holographic principle,
 it is natural to think that  the mass of the test particle $m$
is converted to the horizon energy $\Delta E_h$.
In our theory $E_h$ is simply the total energy inside the horizon.
Therefore,  the following relations
\beqa
mc^2=\Delta E_h
=T_U \Delta S_h=\frac{\hbar a}{2\pi c} \Delta S_h
\eeqa
should hold, which implies for $a=c^2/\Delta x$
\beq
\label{dS}
 \Delta S_h=\frac{2\pi c  m \Delta x}{\hbar},
\eeq
i.e, Eq. (3.6) of Verlinde's paper.
Similarly, Culetu~\cite{Culetu:2010ua} pointed out the  role of the specific distance $\Delta x$
in Verlinde's formatlism.

There have been   criticisms~\cite{Li:2010cj,Culetu:2010ua,Gao:2010yy,Hossenfelder:2010ih} on
 this  entropy variation formula in Verlinde's original model.
The difficulty disappears in our theory, where we identify the Rindler horizon as Verlinde's holographic screen
and the entropy of the  Rindler horizon $S_h$ as the entropy of the screen $S$.

Then, one can define the holographic entropic force
\beq
\label{F}
 F=\frac{\Delta E_h}{\Delta x}= T_U\frac{\Delta S_h}{\Delta x}=ma,
 \eeq
 which is just  Newton's second law.

In short, from the viewpoint of our theory, Verlinde's holographic screen corresponds to Rindler horizons and its entropy is
associated with the lost phase space information of the particle crossing  Rindler horizons~\cite{Lee:2010xv}.
Then, there is an entropic force  linked to this information loss  which can be calculated.
Thus, our theory reproduces and supports Verlinde's mathematical formalism basically.
However, there are several  differences between Verlinde's  model and
our theory, which will be shown in the next section.
Interestingly, this new interpretation seems to also give a hint for the origin of inertia and mass.
The inertia of the particle
can be interpreted as resistance from the horizon dragging which the external force feels.
This dragging force is proportional to acceleration, hence, $F=ma$.

We  see that  inertia and  Newton's second law have something to do with Rindler horizons
 and information loss at the horizons. In our formalism and Verlinde's formalism,
 inertial mass and gravitational mass have a common origin and hence equivalent.
 This is consistent with  the  Einstein's equivalence principle.

 \section{Einstein Gravity from information loss}

 Similarly, one can  interpret Jacobson's formalism and Verlinde's entropic
 gravity in terms of information at Rindler horizons~\cite{Lee:2010xv}.
The equivalence principle allows us to
 choose an approximately flat patch for each  spacetime point.
 According to the principle one can not locally distinguish the free falling frame
 from a rest frame without gravity.
 Therefore,
we can again imagine an accelerating observer $\Theta$ with acceleration $-a$ respect to the test particle
in the rest frame of the particle.
If $\Delta x=c^2/a$, the test particle is just at the   Rindler horizon for the observer $\Theta$,
and there should be energy related to entropy change, i.e., $dE_h=TdS_h$.

Following Jacobson
we can  generalize this information-energy  relation  by defining the
energy flow across the horizon $\Sigma$
\beq
\label{generalized}
dE=-\kappa \lambda\int_\Sigma T_{\alpha\beta} \xi^\alpha d \Sigma^\beta
\eeq
where $d\Sigma^\beta=\xi^\beta d\lambda dA$, $dA$ is the spatial area element,
and $T_{\alpha\beta}$ is the energy momentum tensor of  matter distribution.
Using the Raychaudhuri equation one can
denote the horizon area expansion $\delta A\propto dS_h$  and the increase of the entropy
as
\beq
\label{dA}
dS_h=\eta \delta A=-\eta \lambda \int_\Sigma R_{\alpha\beta} \xi^\alpha  d\Sigma^\beta,
\eeq
with some constant $\eta$ ~\cite{Jacobson}.
If $S_h$ saturates the Bekenstein bound,  $\eta=c^3/4\hbar G$.

Inserting Eqs. (\ref{generalized}) and (\ref{dA}) into $ dE = T_U dS_h=\hbar \kappa dS_h/2\pi c$ one can see
$2\pi c T_{\alpha\beta} \xi^\alpha d \Sigma^\beta
=\eta  R_{\alpha\beta} \xi^\alpha d \Sigma^\beta$. For all local Rindler horizons this equation
should hold. Then, this condition and Bianchi indentity lead to
the Einstein equation
\beq
\label{einstein}
R_{\alpha\beta}-\frac{R g_{\alpha\beta}}{2}+\Lambda g_{\alpha\beta}
=\frac{2\pi }{\eta c} T_{\alpha\beta }
\eeq
with the cosmological constant $\Lambda$ as shown in Jacobson's paper.

\begin{figure}[tpbh]
\includegraphics[width=0.3\textwidth]{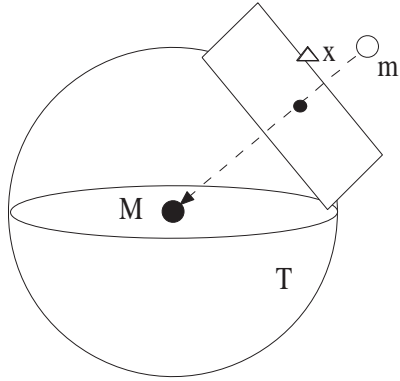}
\caption{
A test particle with mass $m$
is free falling with acceleration $a$ at distance $r$  from a massive object with mass $M$ at the center.
Consider an equivalent situation where is an accelerating  observer $\Theta$
with acceleration $-a$.
If the observer is instantaneously at the distance $\Delta x=c^2/a$
from the test particle, the observer could see the particle crossing
the  local Rindler horizon (the dashed line) for the observer.
}
\end{figure}

Of course, one can derive Newton's gravity from the above Einstein equation.
Alternatively, it is also meaningful to derive Newton's gravity   from $dE=TdS$ relation
and to show that our approach fills the gap between Jacobson's formalism and Verlinde's entropic gravity.

Consider an observer  instantaneously at the distance $\Delta x$ from the test particle with mass $m$.
We can consider a set of such observers surrounding the central mass $M$ at the distance $r+\Delta x$ from the center.
I suggested that the holographic screen considered by Verlinde can be interpreted to be an imaginary overlap of
these local Rindler horizons with a same Unruh temperature $ T_U$ for the observers (Fig. 4).
Again, the mass of the test particle $m$
should be converted to the horizon energy $E_h$
 and this induces the increase of the horizon entropy $\Delta S_h$ eventually.
Therefore,   one can see relations
\beqa
mc^2=\Delta E_h
=T_U \Delta S_h=\frac{\hbar a}{2\pi c } \Delta S_h.
\eeqa
Using the relation $\Delta x=c^2/a$ above, one can obtain the entropy change in Eq. (\ref{dS}) again.
Inspired by the holographic principle  we assume that mass inside a region
is equal to the horizon energy, that is,
\beq
Mc^2=E_h=2T_h S_h=2  T_U S_{BH},
\eeq
where the horizon energy relation $E_h=2T_h S_h$ ~\cite{Padmanabhan:2003pk} and the Bekenstein bound
(i.e., $S_h= S_{BH}$) were used.
The Bekenstein-Hawking entropy
\beq
\label{SBH}
S_{BH}=\frac{ c^3~A}{4G\hbar}
\eeq
 is a bound of  information in a region of space with
 a surface area $A$~\cite{Bekenstein}.
 Since it was shown that the entropy of a Rindler horizon
  is equal to one quarter the area of the  horizon in Planck  units,  this choice is reasonable.
From this equation one can obtain $T_U=Mc^2/2  S_{BH}$ and  the acceleration
\beq
a=\frac{2\pi  c T_U}{\hbar}=\frac{GM}{r^2}.
\eeq
Then, from Eq. (\ref{dS}) and the above equation,  the entropic force is given by
 \beq
 F= T_U\frac{\Delta S}{\Delta x}=\frac{GMm}{r^2},
 \eeq
which is just Newton's gravity.

Therefore, we conclude that the holographic screens at a given position in Verlinde's formalism are
 actually Rindler horizons at the position for specific  observers accelerating relative to the test particle.
This identification could easily explain many questions on Verlinde's formalism
 and provide better grounds for the theory.
 The entropy-distance relation holds only for   specific observers,
 and the use of  Unruh temperature is  valid, because the holographic screen  can be actually
 a set of  Rindler horizons.
 This shows the interesting connection between
 Jacobson's model~\cite{Jacobson} or the  quantum information theoretic model~\cite{Lee:2010bg,Lee:2010fg}
 to Verlinde's model.

However, there are  also several distinctions between Verlinde's  entropic gravity and
our information theoretic model.
First, in Verlinde's work, the screen bounds the emerged part of space, and
 the approaching particle eventually merges with the microscopic degrees of freedom on the screen.
In our theory,  spacetime is not necessarily emergent and the particle
just crosses the horizon and entropy is related to the phase space information loss.
Second, in his theory, the entropy of the screen changes  as the particle approaches
to the screen,
and the screen should move appropriately to satisfy the entropy formula,
while in our theory the change is  due to information loss at  Rindler horizons of specific observers.
Third, in Verlinde's theory the
holographic screens correspond to $equipotential$ surfaces, while
in our theory they correspond to $isothermal$ Rindler horizons
(i.e., with the same $|a|$).
Finally, since Rindler horizons are observer dependent, there is no
objective or observer-independent notion of the Rindler horizon entropy increase in our theory.
This help us to avoid the issue of the time reversal symmetry breaking in entropic gravity.
These differences help us to resolve the possible difficulties
 of Verlinde's original model~\cite{Gao:2010fw}
and to understand the connection between gravity and information.
Compared to other models, our theory emphasizes the role of information rather than thermodynamics.

 \section{Holography and Entanglement  from information loss}
In this section, I explain how the holographic principle and quantum entanglement can arise from the information loss.
The information theoretic derivation of quantum mechanics  in the section III
makes it simple
to understand the information theoretic origin of  the holographic principle~\cite{myholography}.
Consider a $d+1$-dimensional bulk region $\Omega$ with a  $d$-dimensional boundary $\partial \Omega$ that
is an one-way
causal horizon (see Fig. 5).
An outside observer $\Theta_O$   can not access
the information about matter or spacetime in the region
 due to the horizon.
  The best  the observer could do
is to estimate the probability of each possible field configuration
 of $\phi$ in $\Omega$, which turns out to be the probability amplitude in the path integral.
During this estimation, $\Theta_O$ would use  the
maximal information available to her/him.

\begin{figure}[tpbh]
\includegraphics[width=0.4\textwidth]{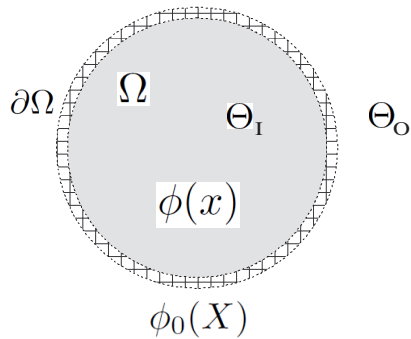}
\caption{A bulk $\Omega$ with a causal horizon $\partial \Omega$ and an inside observer $\Theta_I$.
The outside observer $\Theta_O$ has no   information about the phase space of  $\phi(x)$ in $\Omega$
except for its boundary values $\phi_0(X)$ and derivatives. Thus, according to our information theoretic interpretation,
 the physics in $\Omega$ is completely described
by the boundary physics on $\partial \Omega$, which is just the holographic principle.}
\end{figure}

According to the postulates, there is no non-local  interaction that  might allow  super-luminal communication.
Therefore, we restrict ourselves to  local field theory.
For a local field, any influence on $\Omega$ from the outside of the horizon
should pass the horizon.
 This means that  all the physics in the bulk $\Omega$ is fully described by
 the degrees of the freedom (DOF) on the boundary $\partial \Omega$,
 which is just the essence of the holographic principle!
 Information loss due to a horizon allows the outside observer  $\Theta_O$ to describe the physics in the bulk
 using only the DOF on the boundary. That is the best  $\Theta_O$ can do by any means, and the general
 equivalence principle demands that this description is sufficient  for understanding
 the physics in the bulk,
 which is equivalent to the holographic principle.

Therefore, the following version of  the holographic principle is a natural consequence of the information theoretic formalism of QFT based on   our  postulates.
\begin{theorem}[holographic principle]
 For local field theory, physics inside an 1-way  causal horizon can be  described
 completely by physics on the horizon.
\end{theorem}

Note that this derivation is  generic, because the arguments we used in this section  rely
on neither the specific form of the metric nor any symmetries the fields may have.

Now, how can entanglement  arise in this theory ~\cite{myentanglement}?
Let us assume that the bulk has $N_B$ bits while the surface has $N_b$ bits.
According to the holographic principle,  the bulk
has only area-proportional DOF and hence  there should be redundancy in the bulk bits $B_\alpha$.
Therefore, they are not independent of each other.
We can not simply ignore some of the bulk bits, because
the boundary bits should be able to reproduce arbitrary configuration of the bulk bits,
at least probabilistically. Therefore, only possible way seems to be $n$ to $1$ correspondence between the bulk bits and the boundary bits.
Mathematically, this could mean that there
is a $2^{N_B}$ to $2^{N_b}$  mapping
$f:2^{N_B}\rightarrow 2^{N_b}$.
Since the boundary bits should fully describe the bulk bits (at least probabilistically),
this mapping should be a surjective function.

As a toy example, consider a combination of two bulk bits $B_1$ and $B_2$ which is described by
a single common boundary bit $b_0$ such that
 both of $(B_1,B_2)=(0,0)$ and $(B_1,B_2)=(1,1)$ correspond to $b_0=0$
and both of $(B_1,B_2)=(1,0)$ and $(B_1,B_2)=(0,1)$ correspond to $b_0=1$. Some information in the bulk bits is lost during the mapping.
(This reminds us of the information loss process considered by 't Hooft
in the quantum determinism proposal~\cite{hooft-2002}.
He  introduced equivalence classes of states
that  evolve into one and the same
state.)
Now, assume that $b_0=1$.
This specific  mapping  can be represented by a matrix relation
\begin{eqnarray}
           \begin{bmatrix}
            0  \\
            1  \\
            1  \\
            0
           \end{bmatrix}_B
           =
             \begin{bmatrix}
            1 & 0  \\
            0 & 1  \\
            0 & 1  \\
            1 & 0
           \end{bmatrix}
           \begin{bmatrix}
            0   \\
            1
           \end{bmatrix}_{b},
\end{eqnarray}
where the vector on the left represents the bulk bits in the basis $(00,01,10,11)$
and the vector on the right represents the boundary bits in the basis $(0,1)$, respectively.
The 4 by 2 matrix represents $f$.
With only $b_0$ value
the outside observer can not distinguish two cases
$(B_1,B_2)=(1,0)$ and $(B_1,B_2)=(0,1)$.
Thus, the statistical probability of $b_0$ estimated by the outside observer should be an addition
of two probabilities,
\beq
P_b=P_B((1,0))+P_B((0,1)),
\eeq
where $P_B((1,0))=1/2=P_B((0,1))$ is the probability that  $(B_1,B_2)=(1,0)$
and $P_b=1$ is the probability that $b_0=1$.
In  the path integral formalism derived previously,
for the inside observer this probability  corresponds to an entangled quantum state
\beq
\psi=\frac{1}{\sqrt{2}}(|1\rangle|0\rangle+|0\rangle|1\rangle).
\eeq
Therefore, quantum entanglement is a natural consequence of the holographic principle.
In the information theoretic formalism described in the section III, a quantum state in the bulk corresponds to
a statistical probability like $P_B$ estimated by the outside observer who sees the causal horizon.
This formalism could explain the correspondence between $P_b$ and $\psi$.

 \section{Dark energy from information loss}
Before Verlinde's  proposal  LLK suggested an  idea that
dark energy is related to information content
 of the cosmic horizon~\cite{myDE,Kim:2007vx,Kim:2008re,Lee:2010bg,kias}.
If the cosmic
causal horizon has a radius $r\sim O(H^{-1})$, Hawking  temperature $T\sim 1/r$,  and  entropy
$S\sim r^2$, there could
be
a kind of  thermal
energy $E\sim T S\sim r$
corresponding to the vacuum energy, dubbed `quantum informational dark energy'
by the authors.
Here $H=da/adt$ is the Hubble parameter with the scale factor $a$
and the cosmic horizon could be the event horizon, the Hubble horizon or the apparent horizon.
(There appeared similar dark energy models  based on
the Verlinde's entropic gravity~\cite{Li:2010cj,Zhang:2010hi,Wei:2010ww,Easson:2010av}.)

\begin{figure}[hbtp]
\includegraphics[width=0.32\textwidth]{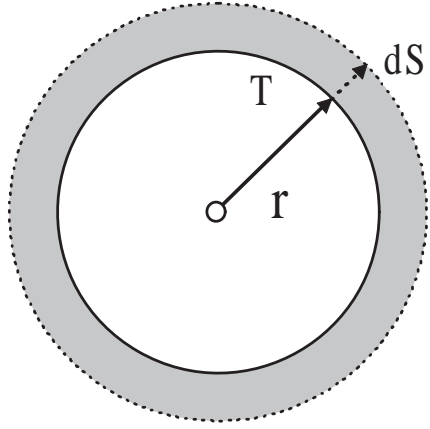}
\caption{
Expansion of a cosmic  horizon $\Sigma$ with a radius $r$ and
the Hawking temperature $T$
induces the information erasing of the gray  region with entropy
 $dS$.
This information erasing consumes the energy  $TdS$, which
 can turn into dark energy finally.
 }
\end{figure}

To calculate the horizon energy $E_h$ as vacuum energy of the universe, let us consider a generic holographic entropy
for a causal cosmic horizon with radius $r$,
\beq
\label{Sh}
S = \frac{\eta  c^3 r^2}{G \hbar},
\eeq
and
\beq
\label{Th}
T=\frac{\epsilon \hbar c}{ r},
\eeq
with parameters $\eta$ and $\epsilon$.
For the Hawking-Gibbons temperature $\epsilon=1/2\pi$, and for the Bekenstein entropy $\eta=\pi$.

Now, one can calculate the vacuum energy using the holographic principle.
By integrating $dE$ on the isothermal surface $\Sigma$
of the causal horizon with Eqs. (\ref{Sh}) and (\ref{Th}), we obtain
the horizon energy
 \beq
 \label{Eh}
E=\int_{\Sigma}  dE=  T \int_{\Sigma}  dS=\frac{\eta \epsilon c^4 r  }{G}.
\eeq
Another possible interpretation is that this is the energy of the cosmic Hawking radiation~\cite{Lee:2008vn}.
Then, the
energy density due to $E_h$  is given by
\beq
\label{rhoh}
\rho_h=\frac{3 E}{4 \pi r^3}
=\frac{6 \eta \epsilon c^3    M_P^2}{\hbar   r^2} \equiv \frac{3 d^2 c^3 M_P^2  }{ \hbar  r^2 },
\eeq
which  has  the form  of the
holographic dark energy~\cite{li-2004-603}.
This kind of dark energy was also derived in terms of entanglement energy~\cite{myDE}
and quantum entanglement force~\cite{Lee:2010fg}.
 From the above
equation  we  immediately obtain a  formula for the constant
 \beq
\label{d1}
 d=\sqrt{{2\eta\epsilon}},
 \eeq
 which is the key parameter determining the characteristic of holographic dark energy.
 The simplest choice is such that
$S_{h}$ saturates the Bekenstein bound
and $T_h$ is the Hawking-Gibbons temperature
${ \hbar c}/{2\pi  r}$. Then, $\eta\epsilon=1/2$ and $d=1$,
which is favored by observations and theories~\cite{Huang:2004ai,zhang-2007}. Thus,
the holographic principle applied to a cosmic causal horizon naturally leads to the holographic dark energy
with $d=1$  ~\cite{Lee:2010fg}!

From the cosmological energy-momentum conservation equation,
one can obtain
an effective  dark energy pressure ~\cite{li-2004-603}
\beq
\label{p}
p_{DE}=\frac{d(a^3\rho_h(r))}{-3 a^2 da },
\eeq
from which one can derive the equation of state.
To compare  predictions of our theory  with current observational data,
 we need to choose the horizon.
The event horizon is the simplest one, if there is no interaction term between dark energy and
matter~\cite{li-2004-603}. In this case
one can find  the equation of state for holographic dark energy as a function of the redshift $z$ \cite{li-2004-603};
\beqa
\label{omega3}
\omega_{DE} &=&\left (1+\frac{2\sqrt{\Omega^0_\Lambda}}{d}\right)
\left(-\frac{1}{3}
+z\frac{\sqrt{\Omega^0_\Lambda}(1-\Omega^0_\Lambda)}{6d}\right) \no
 &\simeq& w_0+w_1 (1-a),
\eeqa
where the current dark energy density parameter $\Omega^0_\Lambda\simeq 0.73$
~\cite{li-2004-603,1475-7516-2004-08-006}.
For $d=1$ these equations give $w_0=-0.903$ and $w_1=0.208$.
According to
  WMAP 7-year data with
 the  baryon acoustic oscillation, SN Ia, and the Hubble constant
  yields $w_0 = -0.93\pm 0.13$
  and  $w_1 = -0.41^{+0.72}_{-0.71}$ ~\cite{Komatsu:2010fb}.
Thus, the predictions of our theory well agree with the recent observational data.
  If we use an entanglement entropy calculated in \cite{Lee:2010fg} for $S_h$,
one can obtain $d$ slightly different from $1$.

It was also shown that holographic dark energy models  with an inflation with
a number of e-folds $N_e\simeq 65$
can solve the cosmic coincidence problem~\cite{mycoin,li-2004-603}
thanks to a rapid expansion of the event horizon during the inflation.

Following \cite{Lee:2010fg} and \cite{Easson:2010av} one can obtain an entropic force
for the dark energy
\beq
\label{entforce}
F_{h}\equiv \frac{dE_{h}}{dr}=\frac{c^4 \eta\epsilon}{ G},
\eeq
which could be also identified as  a `quantum entanglement force' dubbed by LLK,
if $S_h$ is the entanglement entropy.

It is simple to see why the cosmological constant $\Lambda_c$ should be zero.
The classical cosmological constant $\Lambda_c$  appears in the gravity action as
\beq
\label{a}
S=\int d^4x \sqrt{-g}(R-2\Lambda_c).
\eeq

It is usually argued that after taking vacuum expectation of  quantum fields,
the Friedmann equation
has additional
  contribution $\Lambda_q=\rho_q/M_P^2 c^2$ from the vacuum quantum fluctuation $\rho_{q}$.
Thus, the total cosmological constant becomes
$\Lambda=\Lambda_c +\Lambda_q,$
and
the total vacuum energy density is given by
\beq
\label{rhovac2}
\rho_{vac}= M_P^2 c^2 (\Lambda_c  + \Lambda_q  ).
\eeq
Without a fine tuning it is almost impossible for two terms to cancel
each other to reproduce the tiny observed value, which
 is the well-known  cosmological constant problem.


A constant $\Lambda_c$ results in  vacuum energy
   proportional to $\Lambda  r^3$ clearly violating the holographic principle for large $r$
   (where matter energy density of the universe is small), because
   $E_h\propto r$ according to the principle and the information-energy relation.
This implies that  the `time independent' classical cosmological constant $\Lambda_c$ should be zero
  and  $\Lambda_q$ is  proportional to $\rho_h$ in Eq. (\ref{rhoh}), unless there is interaction
  between matter and dark energy.
Of course, this argument does not show how to remove the cosmological constant explicitly in QFT.
QFT is not one of our assumptions  but derived with  specific conditions.
Since the holographic principle is in contradiction with
QFT at a large scale, this might mean that we need to change QFT at a cosmological scale.

In summary, in this theory
the dark energy density is small due to the holographic principle,
 comparable to the critical density due to the  $O(1/H)$ horizon size
 or   $N_e\simeq 65$,
and non-zero due to  quantum vacuum fluctuation.
 The holographic principle also
 demands that  the cosmological constant is zero.

\section{Discussion}

In short, the Einstein equation links matter to gravity
and his famous formula $E=mc^2$ links matter to energy.
We know also that the Landauer's principle links  information to  energy.
Thus, now we have  relations among information, gravity, quantum mechanics and
classical mechanics.
Our theory implies that physical laws are more about information
rather than particles or waves.
Quantum randomness and its thermal nature arise from information loss
 at causal horizons.
This gives us a new hint of quantum gravity.
 Our new approach also  shows  interesting connections between
 Jacobson's model~\cite{Jacobson}, the  quantum information theoretic model~\cite{Lee:2010bg,Lee:2010fg}
 and Verlinde's model for gravity.

We also see that  inertia and  Newton's second law have something to do with Rindler horizons
 and information loss at the horizons. In our formalism and Verlinde's formalism,
 inertial mass and gravitational mass have a common origin and hence equivalent.

The holographic principle and quantum entanglement can be explained easily in this formalism.
All these studies are not a simple reinterpretation of existing physics.
If information really is the essence of the universe, this alters our very paradigm
in looking at physics, and it may serve as a key to solving hard problems in the field such
as a theory of everything, dark energy, and quantum gravity.

\section*{acknowledgments}
Author is thankful to H. Culetu, Jungjai Lee, Hyeong-Chan Kim, and Gungwon Kang for
helpful discussions.
This work was supported in part by
Basic Science Research Program through the National Research Foundation of Korea (NRF) funded by
the ministry of Education, Science and Technology (2010-0024761)
and the topical research
program (2010-T-1) of Asia Pacific Center for
Theoretical Physics.

%

\end{document}